\documentclass[prd,twocolumn,superscriptaddress,showpacs,nofootinbib,preprintnumbers]{revtex4}
\usepackage{amsmath}
\usepackage{amsfonts}
\usepackage{graphicx}
\usepackage{dcolumn}
\newcommand{\beq}{\begin{equation}}
\newcommand{\eeq}{\end{equation}}
\newcommand{\bea}{\begin{eqnarray}}
\newcommand{\eea}{\end{eqnarray}}
\newcommand{\beaa}{\begin{eqnarray*}}
\newcommand{\eeaa}{\end{eqnarray*}}

\begin{document}
\tolerance=5000


\title{Cosmological investigation of multi-frequency VLBI observations of ultra-compact structure in $z\sim 3$ radio quasars}

\author{Shuo Cao}
\affiliation{Department of Astronomy, Beijing Normal University,
Beijing 100875, China; zhuzh@bnu.edu.cn}

\author{Marek Biesiada}
\affiliation{Department of Astrophysics and Cosmology, Institute of
Physics, University of Silesia, Uniwersytecka 4, 40-007, Katowice,
Poland }

\author{Jingzhao Qi}
\affiliation{Department of Physics, College of Sciences,
Northeastern University, Shenyang 110004, China}

\author{Yu Pan}
\affiliation{College of Science, Chongqing University of Posts and
Telecommunications, Chongqing 400065, China}

\author{Xiaogang Zheng}
\affiliation{School of Physics and Technology, Wuhan University,
430072, Wuhan, China}

\author{Tengpeng Xu}
\affiliation{Department of Astronomy, Beijing Normal University,
Beijing 100875, China; zhuzh@bnu.edu.cn}

\author{Xuan Ji}
\affiliation{Department of Astronomy, Beijing Normal University,
Beijing 100875, China; zhuzh@bnu.edu.cn}

\author{Zong-Hong Zhu$^{\ast}$}
\affiliation{Department of Astronomy, Beijing Normal University,
Beijing 100875, China; zhuzh@bnu.edu.cn}

\begin{abstract}

In this paper, we use multi-frequency angular size measurements of
58 intermediate-luminosity quasars reaching the redshifts $z\sim 3$
and demonstrate that they can be used as standard rulers for
cosmological inference. Our results indicate that, for the majority
of radio-sources in our sample their angular sizes are inversely
proportional to the observing frequency. From the physical point of
view it means that opacity of the jet is governed by pure
synchrotron self-absorption, i.e. external absorption does not play
any significant role in the observed angular sizes at least up to 43
GHz. Therefore, we use the value of the intrinsic metric size of
compact milliarcsecond radio quasars derived in a cosmology
independent manner from survey conducted at 2 GHz and rescale it
properly according to predictions of the conical jet model. This
approach turns out to work well and produce quite stringent
constraints on the matter density parameter $\Omega_m$ in the flat
$\Lambda$CDM model and Dvali-Gabadadze-Porrati braneworld model. The
results presented in this paper pave the way for the follow up
engaging multi-frequency VLBI observations of more compact radio
quasars with higher sensitivity and angular resolution.

\end{abstract}


\maketitle

\section{Introduction}\label{sec:introduction}


For some time, radio sources (extended FRIIb radio galaxies, radio
loud quasars, etc.) have been proposed as standard rulers
\citep{Buchalter98,Guerra98,Guerra00,Daly03} and hence as
alternative cosmological probes complementary to standard candles
(SN Ia) and anisotropies in the cosmic microwave background
radiation (CMBR). One important class of such objects are
ultra-compact radio-sources whose cores have angular sizes of order
of milliarcseconds ($mas$) which could be measured by
very-long-baseline interferometry (VLBI)
\citep{Kellermann93,Gurvits94}. In the VLBI images, the core is
usually identified with the most compact (often unresolved) feature
with a substantial flux and flat spectrum across the radio band.
More importantly, radio sources (especially quasars) can be observed
up to very high redshifts, well beyond the observed redshift range
of SNIa \citep{Amanullah10} limited to $z<2$. First attempts to
constrain cosmological models using such kind of sources were due to
\citet{Gurvits99}, who compiled a data-set of 330 milliarcsecond
radio sources containing various optical counterparts (radio
galaxies, quasars, BL Lac etc.).
A sub-sample containing 145 compact sources with little dependence
of angular size on spectral index ($-0.38\leq \alpha\leq 0.18$) and
luminosity $Lh^2 \geq 10^{26} \; WHz^{-1}$ was also derived in their
analysis, based on which all data points were distributed into
twelve redshift bins and were extensively discussed in the
literature \citep{Vishwakarma01,Zhu02,Chen03} as cosmological
probes. However, the determination of the typical value of the
linear size $l_m$ for this standard ruler (or even whether compact
radio sources are indeed ``true'' standard rulers) was remaining an
important problem to be solved \citep{Vishwakarma01,Cao15}. Based on
the previous conclusion formulated in \citet{Cao15}, that mixed
population of radio sources including different optical counterparts
can not be treated as a ``true'' standard ruler, \citet{Cao16}
demonstrated that only in the intermediate-luminosity radio quasars
their compact structure displayed minimal dependence of the linear
size on redshift and luminosity. Therefore they could serve as a
cosmological standard ruler. Identifying 120 such
intermediate-luminosity quasars allowed \citet{Cao17a} to explore
various cosmological models in the high redshift range ($z\sim3.0$),
a range which was difficult to access by other cosmological probes.
Such quasar sample has also been extensively used to investigate
other dynamical dark energy models \citep{Li17,Zheng17} and modified
gravity theories \citep{Qi17,Xu17}. This was done however, based on
the results of a survey performed on a single frequency 2.29 GHz and
a question might be raised whether this successful calibration of
standard rulers is just by coincidence or has a broader scope of
applications comprising VLBI surveys on other frequencies. The focus
of this paper is to extend the previous analysis of \citet{Cao17a}
and investigate possible astrophysical applications of the
multi-frequency angular-size measurements of 58 quasars covering
redshifts $z = 0.536 - 2.73$. More importantly, the angular size of
ultra-compact structure depends on the observing frequency, due to
synchrotron self-absorption of the radio core (absorption in the
radio emitting plasma itself) and external absorption in the
surrounding material \citep{Blandford79,Lobanov98}. Therefore, the
measurements of the angular sizes at three or more frequencies can
be used to study the physics of compact radio-emitting region.

\begin{table*}
\caption{\label{tab:data} Compilation of intermediate-luminosity
quasars from \citet{Pushkarev15}. Quasars with flux density smaller
than 1.0 Jy are written in bold. Column (1): source (name); Column
(2): redshift; Column (3)-(8): angular size in milliarcseconds at 2,
5, 8, 15, 24, and 43 GHz, respectively.}

\begin{tabular}{llllllll|llllllllll}

\hline\hline

Source & $z$ & $\theta_2$ & $\theta_5$ & $\theta_8$ & $\theta_{15}$
& $\theta_{24}$ & $\theta_{43}$ & Source & $z$ & $\theta_2$ &
$\theta_5$ & $\theta_8$ & $\theta_{15}$
& $\theta_{24}$ & $\theta_{43}$ \\

\hline

J1256-0547 & 0.536 &  2.56  &  0.59  &  0.37  &  0.23 &   0.13 & 0.07   & \textbf{J1617+0246} & 1.339 & 1.91 &&  0.33 &&&\\
J0407-1211 & 0.573 &  1.82  & &  0.33 & & &   & \textbf{J0808+4052} & 1.42 & 0.83 & 0.43 & 0.15 & 0.06 & 0.12 & 0.07 \\
J0922-3959 & 0.591 &  2 & & 0.38 &&&   &J1534+0131 & 1.435 & 1.24 & 0.86 & 0.43 & 0.18 &&\\
J1642+3948 &  0.593 &  1.29 &   1.28 &   0.54  &  0.24  &  0.19 &  &\textbf{J1033-3601} & 1.455 & 1.08 && 0.48 &&&\\
J2332-4118 & 0.671 &  1.89 &&   0.45 &&&   &J2255+4202 & 1.476 & 0.99 & 0.52 & 0.39 &&& \\
J1800+7828 & 0.68 &   0.55 & 0.27 &0.23& 0.16& 0.1& 0.09    &J2056-4714 & 1.489 & 2 &&  0.59 &&&\\
J1357+1919 & 0.72 &   1.4 & 0.75 &0.41 &   0.14  &  0.43 &  &\textbf{J0222-3441} & 1.49 & 0.98 && 0.45 &&& \\
J1637+4717 & 0.74 & 0.72  &&  0.23  &  0.08 &&    &\textbf{J1417+4607} & 1.554 & 1.7 & 1.63 & 0.74 &&&\\
J1239-1023 & 0.752 & 2.25 & 1.37 & 0.7 &&&    &J2229-0832 & 1.56 & 0.82 && 0.18 & 0.11 & 0.07 & 0.04   \\
J0728+6748 & 0.846 &  0.99 &&  0.26 &&&    &\textbf{J0409-1238} & 1.563 & 0.81 &&  0.22 &&& \\
\textbf{J0917-2131} & 0.847 & 1.72  && 0.2 &&&      &\textbf{J0839+0319} & 1.57 & 1.7 && 0.49 &&&\\
J1215+3448 & 0.857 & 1.31  & 0.82  & 0.51 &&&     &J1107-4449 & 1.598 &  2.84 && 0.9 &&&\\
J0538-4405 & 0.894 & 1.41 &&  0.43 &&&     &J1640+3946 & 1.66 & 0.64 & 0.28 & 0.27 & 0.12 & 0.08  &  0.05   \\
\textbf{J0539-1550} & 0.947 &  1.59  &&   0.38 &&&    &J0110-0741 &1.776 & 1.85  &&&&&\\
J1937-3958 & 0.965 & 1.49  &&   0.34 &&&     &\textbf{J1454-4012} & 1.81 & 2.17 && 0.46 &&&\\
\textbf{J0239+0416} & 0.978 &  0.89 && 0.23 & 0.05 &&   &\textbf{J1036-3744} & 1.821 & 1.37 &&&&& \\
\textbf{J0132-1654} & 1.02 & 1.44 & 0.59 & 0.91 & 0.25 &&    &J0808-0751 & 1.837 &  1.18 & 1.23 & 0.29 & 0.1 & 0.07 &\\
\textbf{J1516+1932} & 1.07  &  0.88 && 0.16 & 0.09 &   0.11 &   &\textbf{J0639+7324} & 1.85 & 1.44 &&  0.26 &&& \\
\textbf{J1337+5501} & 1.1 & 1.22 & 0.45 &   0.47 &&&    &J1357-1527 & 1.89 & 0.96 &&  0.23 & 0.14 & 0.12 & 0.07  \\
\textbf{J1213+1307} & 1.14 & 1.53 &&&&&   &\textbf{J0620-2515} & 1.9 & 0.84 &&  1.09 &&&\\
J2331-1556 & 1.153& 1.56 & 1.15 & 0.5 & 0.45 &&    &\textbf{J1658+3443} & 1.937 & 1.82 & 0.47 & 0.57 &&&\\
\textbf{J1441-3456} & 1.159 &  1.9 &&&&&    &\textbf{J1327+4326} & 2.08 & 0.94 & 0.41 & 0.57 &&&\\
J1955+5131 & 1.22 & 1.05 & 0.66 & 0.23 & 0.2 &&    &\textbf{J2322+0812} & 2.09 & 2.06 && 1.14 &&&\\
\textbf{J1153+8058} & 1.25 & 1.19 &&  0.31 & 0.27 & 0.15 &    &\textbf{J1022+1853} & 2.136 & 1.94 & 2.39 &&&& \\
J1023+3948 & 1.254 & 1.03 & 0.69 & 0.26 & 0.1 &&     &J0644-3459 & 2.165 & 3.28 & 0.57 & 0.91 &&&\\
\textbf{J0516-1603} & 1.278 & 1.69 &&   0.73 &&&    &J1035-2011 & 2.198 &  1.81 & 0.89 & 0.5 & 0.41 &   0.21 &\\
J0406-3826 & 1.285 & 0.96  && 0.45 &&&     &\textbf{J2316-4041} & 2.448  & 1.19 &&   0.21 &&&\\
\textbf{J0710+4732} & 1.292 &  0.79 & 0.75 & 0.16 & 0.12 & 0.07 &   &\textbf{J0331-2524}  & 2.685 & 1.77 && 0.2 &&&\\
\textbf{J2314-3138} & 1.323 & 0.91 &&  0.36 &&&        &\textbf{J0139+1753} & 2.73 & 1 && 0.72 & 0.23 &&\\

\hline\hline

\end{tabular}

\end{table*}

\section{Methodology and observational Data}

Standard ruler approach to measure cosmological distances
\citep{Sandage88}, is based on quite obvious geometric relation
\begin{equation}
\theta(z)= \frac{l_m}{D_A(z)} \label{theta}
\end{equation}
between the intrinsic metric length $l_m$ of the standard ruler
located at the redshift $z$, its observed angular size $\theta(z)$
and its angular diameter distance $D_A(z)$. The main problem here is
to find a convincing population of standardizable rulers. In
particular, metric sizes $l_m$ of compact radio sources may depend
on their luminosity $L$ (i.e. on the central engine) and display
evolutionary effects, i.e. may depend on $z$. As already mentioned,
\citet{Cao16} using the parametrization $l_m=lL^\beta(1+z)^n$
capturing these effects, demonstrated that the linear size $l_m$ of
compact structure in 120 intermediate-luminosity quasars observed at
2.29 GHz (later on we will denote this frequency as 2 GHz for short)
displays negligible dependence both on redshift and luminosity
($|n|\simeq 10^{-3}$, $|\beta|\simeq 10^{-4}$).

In extragalactic jets, however, the apparent position of a bright
narrow end depends on the observing frequency, owing to synchrotron
self-absorption and external absorption. Considering that 10 pc is a
typical radius at which AGN jets are apparently generated
\citep{Blandford78}, it is very important to investigate the
relation between the observing frequency and the apparent linear
size of compact structure. To be more specific, at any given
frequency, the core is believed to be located in the region of the
jet where the optical depth is $\tau=1$. In the conical jet model
proposed by \citet{Blandford79}, if we observe a given
milliarcsecond source at different observing frequencies $\nu$, its
observed size falls as the frequency increases
\citep{Marscher80,O'Sullivan09}, being proportional to $\nu^{-1}$.
At this point, one should clarify the issue of reception frequency
$\nu_r$ and the rest-frame frequency $\nu_e$. At first sight an
expectation would be that angular sizes in corresponding
angular-size vs. redshift diagrams should fall by an additional
factor $(1+z)^{-1}$, because the rest-frame emitted frequency has to
be greater than $\nu_e$ by a factor of $(1+z)$: $\nu_e=(1+z)\nu_r$,
thus masking any cosmological effect. However, observations show
behavior which is compatible with conventional cosmologies. The
reason is almost certainly that there is a selection effect which is
operating in our favor in this context. Namely, we are dealing with
an ensemble of objects which may be intrinsically similar in their
respective rest-frames, but appear to be very different in our
frame. According to the unified model of active galactic nuclei and
quasars \citep{Antonucci93, Antonucci15}, the underlying population
consists of compact symmetric objects \citep{Wilkinson94}, each
comprising a central low-luminosity nucleus straddled by two
oppositely-directed jets. Ultra-compact objects are identified as
cases in which the jets are moving relativistically and are close to
the line of sight, when Doppler boosting allows just that component
which is moving towards the observer to be observed. Giving rise to
the core-jet structure observed in typical VLBI images
\citep{Pushkarev12}, the core tends to act as the base of the jet
instead of the nucleus \citep{Blandford79}. Those jets which are
closest to the line of sight appear to be the brightest. Following
the analysis of \citet{Dabrowski95}, for a flux-limited radio
sample, a larger Doppler boost factor {\cal D} is required as $z$
increases, which will generate an approximately fixed ratio ${\cal
D}/(1 + z)$ and thus an approximately fixed rest-frame emitted
frequency $(1 + z)\nu_r/{\cal D}$. See \citet{Jackson04} for
mathematical and astrophysical details.

Moreover, the dependence of the angular size ($\theta$) on spectral
index ($\alpha$) which, if not considered, could also constitute a
possible sources of systematics in the dispersion of the linear size
($l_m$). More specifically, following the simple consideration of
self absorption arguments, radio sources with flat and inverted
spectra tend to have smaller sizes and there is an obvious
dependence of angular size on spectral index (see Fig.~7 of
\citet{Gurvits99} for details). In view of this effect in the
currently available sample, there have been arguments based on the
restricted range of spectral indices (a flat segment of the $\theta
- \alpha$ diagram $-0.38 \leq \alpha \leq 0.18$) that elimination of
the large compact steep spectrum sources and most compact inverted
spectrum sources will better define compact sources as standard
rulers \citep{Gurvits99}. More importantly, as was noted in the same
work, the remarkable feature of such selection criterion lies in the
fact that lowest-redshift sources ($z<0.5$), which exhibit the
highest deviation on the $\alpha - z$ diagram \citep{Jackson04},
will be partially excluded from the final sample used for
statistical analysis. Therefore, in this work we will adopt the same
criterion of $\alpha$ and concentrate on compact radio quasars with
flat spectral index.

\begin{figure}
\begin{center}
  \includegraphics[scale=0.45]{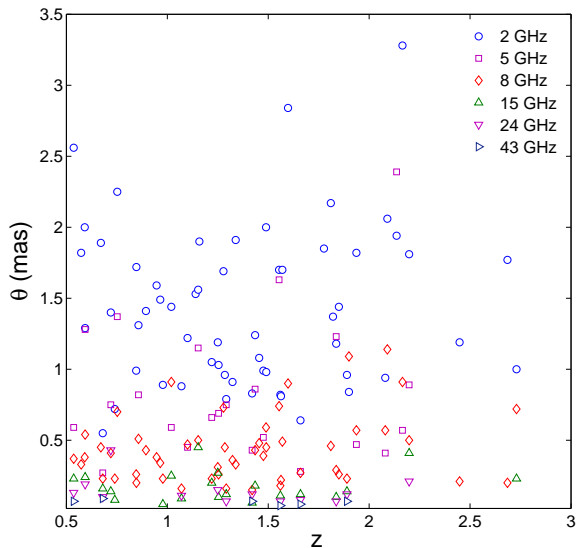}
  \end{center}
  \caption{ Angular size vs. redshift for the \citet{Pushkarev15} sample of 58 sources at six radio frequencies. }\label{figP15}
\end{figure}

In our analysis, we turn to the more recent VLBI imaging
observations based on better uv-coverage. \citet{Pushkarev15}
presented the VLBI data of more than 3000 compact extragalactic
radio sources observed at different frequencies, $\nu = 2 - 43$ GHz.
This sample, however, contains a wide class of extragalactic
objects, belonging to different luminosity categories, including
quasars, radio galaxies, and BL Lac objects. We identified 58
intermediate-luminosity quasars from the sub-sample constructed in
\citet{Cao16}, which have also been included in the
\citet{Pushkarev15} sample. Fig.~\ref{figP15} displays the observed
angular size against redshift in this sample. This way, we obtained
a set of data -- summarized in Table~1 -- comprising angular sizes
of flat spectrum cores in intermediate-luminosity radio quasars at
six different radio frequencies: 2, 5, 8, 15, 24 and 43 GHz.

\begin{figure}
\begin{center}
  \includegraphics[scale=0.4]{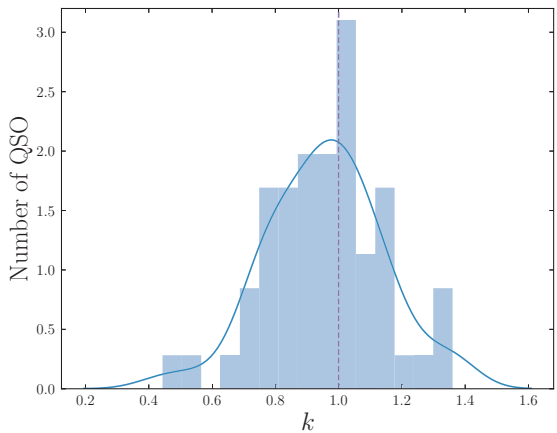}
  \end{center}
\caption{ Distribution of $k$ parameter in our sample. Blue solid
curve represents the best fitted continuous distribution function.
Pure synchrotron self-absorption case ($k=1$) is denoted by the
purple dashed line. }\label{figk}
\end{figure}

In order to use Eq.(\ref{theta}) for cosmological inference one
needs to calibrate $l_m$. According to \citet{Blandford79}
calibrated metric size of such standard ruler should depend on
observing frequency in the same way as the observed angular size.
Owing to synchrotron self-absorption of the radio core (absorption
in the radio emitting plasma itself) and external absorption in the
surrounding material, the angular size at any given observing
frequency $\nu$ should scale as $\theta_{\nu} \propto \nu^{-k}$. The
parameter $k$ representing the dependence of the angular size on
frequency, is related to physics of the compact radio-emitting
region: the shape of electron energy spectrum and the magnetic field
and particle density distributions. Namely, (1) $k=1$ if the core is
self-absorbed and in equipartition \citep{Blandford79}; (2) $k>1$ if
external absorption plays an important role \citep{Lobanov98}.
Therefore, the measurements of the angular sizes at three or more
frequencies can be used for determining the value of $k$ and thus
testing the \citet{Blandford79} jet model and its later
modifications \citep{Hutter86,Bloom96}. Therefore, VLBI results
obtained at other frequencies on the same sources as used in the
current study should be used for verification of the results. We
checked that frequency dependence of angular sizes in our sample is
compatible with \citet{Blandford79} conical jet model. For each
quasar, the best-fit $k$ parameter and its corresponding 1$\sigma$
uncertainty is determined through a $\chi^2$ minimization method.
Following the previous analysis of a large sample of 140 mas compact
radio sources \citep{Cao15}, we have assumed a conservative 10\%
statistical uncertainty in the observed angular sizes. Fig.~2
displays the distribution of $k$ parameter in our sample of quasars.
Best-fit probability distribution function (PDF) is represented by
the blue solid curve.

One objection one might rise towards the proposed method is the
rigid assumption of the \citet{Blandford79} conical jet model.
Although there are some arguments supporting good consistency
between the VLBI observations and the BK79 conical jet model in the
earlier studies, which was done by using the recent multi-frequency
core shift measurements of many compact radio sources
\citep{Lobanov98,Kovalev08,Kovalev09,O'Sullivan09}, one can expect
the deviation from this standard jet model. Therefore, in the
context of multi-frequency data, one can use the characteristic
linear size at 2 GHz and scale it according to $l_m\propto\nu^{-k}$
to any other frequency. However, this model still requires that the
quantity $k$ is constant, while as we see in Fig.~2 it has some
distribution across the sample. Consequently in the following
analysis we also performed fits assuming the linear relation:
$k(z)=k_0+k_1z$ treating $k_0$ and $k_1$ as free parameters together
with cosmological ones. We remark here that, considering the fact
that the fixed ratio ${\cal D}/(1 + z)$ is only expected to occur at
the flux limit of the survey, it is mandatory to take this effect
into consideration and verify its usability for cosmological test.
Therefore, in the following analysis we will also use 30 quasars
with flux density smaller than 1.0 Jy, a flux-density limit
resulting in a sample observed with VLBI covering the full sky
\citep{Preston85}. This restricted sample is summarized in Table 1
where the names of quasars are given in bold.

Now a cosmological-model-independent method should be applied to
derive the linear size of the compact structure in radio quasars at
2 GHz, by constructing angular diameter distances $D_A$ by means of
GP-processed $H(z)$ measurements \citep{morescoetal,Zheng16} from
cosmic chronometers \citep{JimenezLoeb} (using publicly available
GaPP code \citep{Seikel12a}). See \citet{Cao17a} for detailed
description of this procedure. The advantage of our quasar sample,
compared with other reliable standard rulers extensively used in the
literature: baryon acoustic oscillations (BAO)
\citep{Roukema2015,Roukema2016} and galaxy clusters with radio
observations of the Sunyaev-Zeldovich effect and X-ray emission
\citep{Filippis05,Boname06}, lies in the fact that quasars are
observed at much higher redshifts ($z\sim3$). More importantly, the
angular diameter distance information obtained from quasars has
helped us to bridge the "redshift desert" and extend our
investigation of dark energy to much higher redshifts, reaching
beyond feasible limits of supernova studies ($z\sim1.4$)
\citep{Amanullah10}.

Based on the 2 GHz angular-size measurements from the P15 sample, we
estimate the characteristic linear size as $l_{m,2}= 8.76\pm0.25$
pc, which is scaled to other frequency at which angular size was
observed. Note that the calibration result obtained in this  paper
is slightly different from that obtained in the previous work within
$1\sigma$, which used a compilation of 120 milliarcsecond compact
radio-sources representing intermediate-luminosity quasars
\citep{Cao17a}. However, such mild tension will be well resolved if
$2\sigma$ uncertainty is taken into account. There exist several
possible explanations of this possible tension or incompatibility
between the single-frequency and multi-frequency measurements. First
of all, considering the fact that the sources used here are only a
subset of those used previously, it may be a statistical result
produced by the limited amount of observational data. Another
important element producing this discrepancy can be ascribed to the
intrinsic difference of the angular-size measurements given
different techniques for image reconstruction. The data used in the
previous works are derived from an ancient VLBI survey undertaken by
\citet{Preston85}, which defines a characteristic angular size
through the ratio of total flux density and correlated flux density
(fringe amplitude) \citep{Jackson04,Jackson06}. In this analysis, we
have used multi-frequency VLBI observations of more compact radio
quasars with higher sensitivity and angular resolution, in which,
following the approach by \citet{Kovalev05}, the most compact
component assigned as the VLBI core in contour maps is fitted to the
self-calibrated visibility data for all sources \citep{Pushkarev15}.

\begin{table*}
\caption{\label{result} Constraint results obtained by the quasar
sample for the flat $\Lambda$CDM and DGP model.}
\begin{center}
\begin{tabular}{l|lllll}\hline\hline
 Cosmology ($H_0$ Priors) & \hspace{4mm}  $\Omega_m$ &\hspace{4mm} $k_0$ &\hspace{4mm} $k_1$ \hspace{4mm} \\ \hline
$\Lambda$CDM1 (Planck 2014)  & $\Omega_m=0.353\pm0.063$    & $k_0=0.988\pm0.025$  & $k_1=0$  \\
$\Lambda$CDM2 (Planck 2014)  & $\Omega_m=0.348\pm0.063$   & $k_0=0.999\pm0.043$  & $k_1=-0.009\pm0.032$   \\
$\Lambda$CDM1 (Riess 2016)  & $\Omega_m=0.239\pm0.047$    & $k_0=0.995\pm0.023$  & $k_1=0$  \\
$\Lambda$CDM2 (Riess 2016)  & $\Omega_m=0.229\pm0.050$   & $k_0=1.030\pm0.042$  & $k_1=-0.031\pm0.032$   \\
\hline
DGP1 (Planck 2014)  & $\Omega_m=0.252\pm0.055$    & $k_0=0.989\pm0.024$  & $k_1=0$  \\
DGP2 (Planck 2014)  & $\Omega_m=0.245\pm0.054$   & $k_0=1.009\pm0.043$  & $k_1=-0.016\pm0.032$   \\
 \hline\hline
\end{tabular}
\end{center}

\end{table*}

Thus, for the analysis of the multi-frequency quasar data, we
perform Monte Carlo simulations of the posterior likelihood ${\cal
L} \sim \exp{(- \chi^2 / 2)}$, where
\begin{equation}
\chi^{2} =\sum_{j=1}^{6}\sum_{i=1}^{30}
{\frac{\left[\theta^{th}_{i,j}(l_{m,j}; \textbf{p},
k)-\theta^{obs}_{i,j}\right]^{2}}{\sigma_{i,j}^{2}}}.
\end{equation}
where the $k$ parameter is fitted together with cosmological
parameters \textbf{p}. The summation is over different quasars at
redshifts $z_i$ observed at different frequencies $\nu_j$. Note that
the data point with missing frequency, which is not included in the
summation, will not contribute to the $\chi^2$ statistics.
$\theta^{th}_{i,j}=l_{m,j}/D_{A,i}$ is the theoretical value of the
angular size of an object of proper length $l_{m,j}$ at observing
frequency $\nu_j$, while $\theta^{obs}_{i,j}$ is the corresponding
observed value with total uncertainty $\sigma_{i,j}$. Following the
previous work of \citet{Cao17a}, in this analysis the total
uncertainty expresses as
$\sigma^2_{i,j}\,=\,\sigma^2_{sta,i,j}+\sigma^2_{sys,i,j}$. We have
assumed $10\%$ statistical error of observations in
$\theta_{i,j}^{obs}$ and an additional $10\%$ systematical
uncertainty accounting for the intrinsic spread in the linear size.

\begin{figure*}
\begin{center}
\includegraphics[scale=0.45]{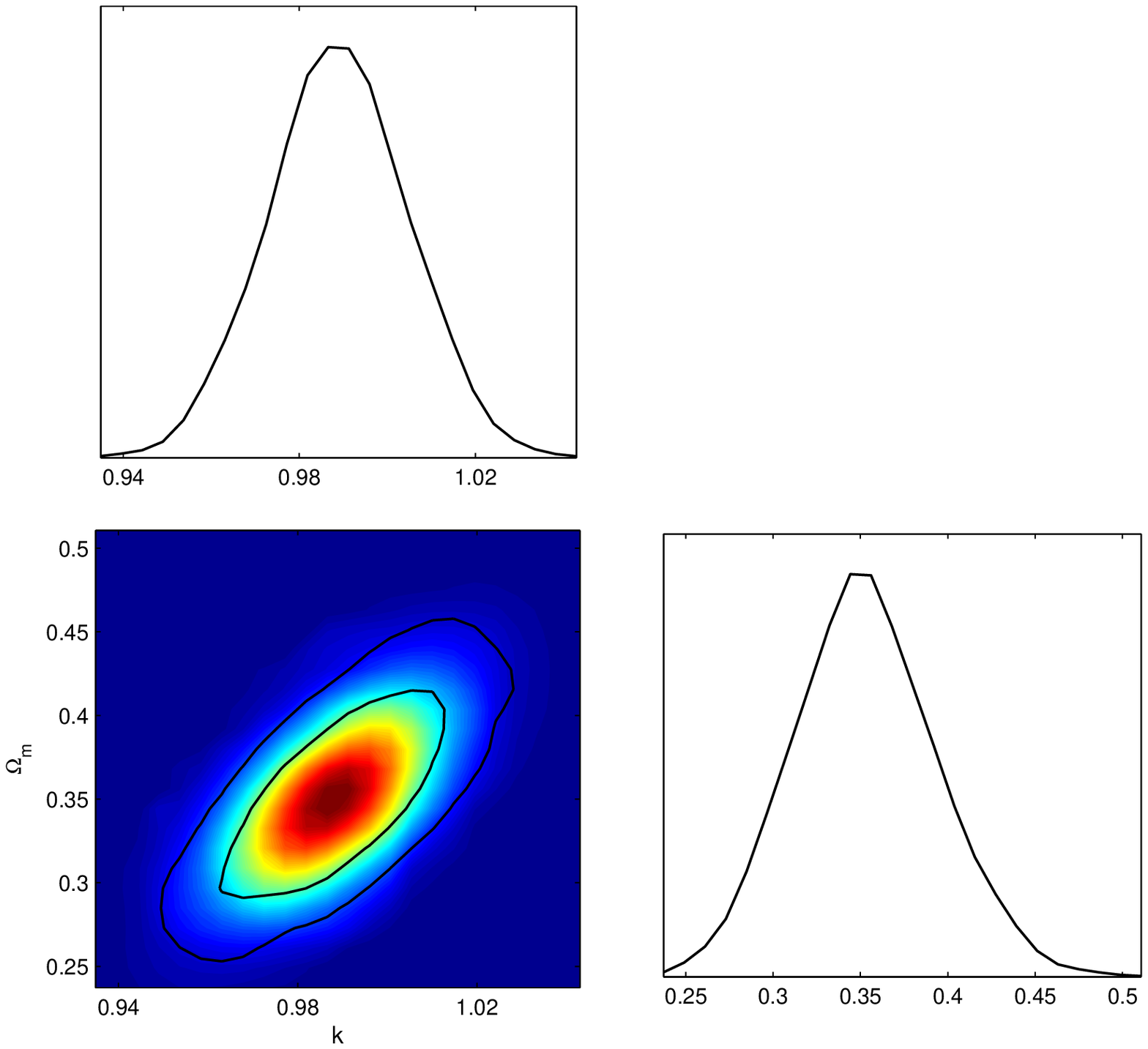} \includegraphics[scale=0.45]{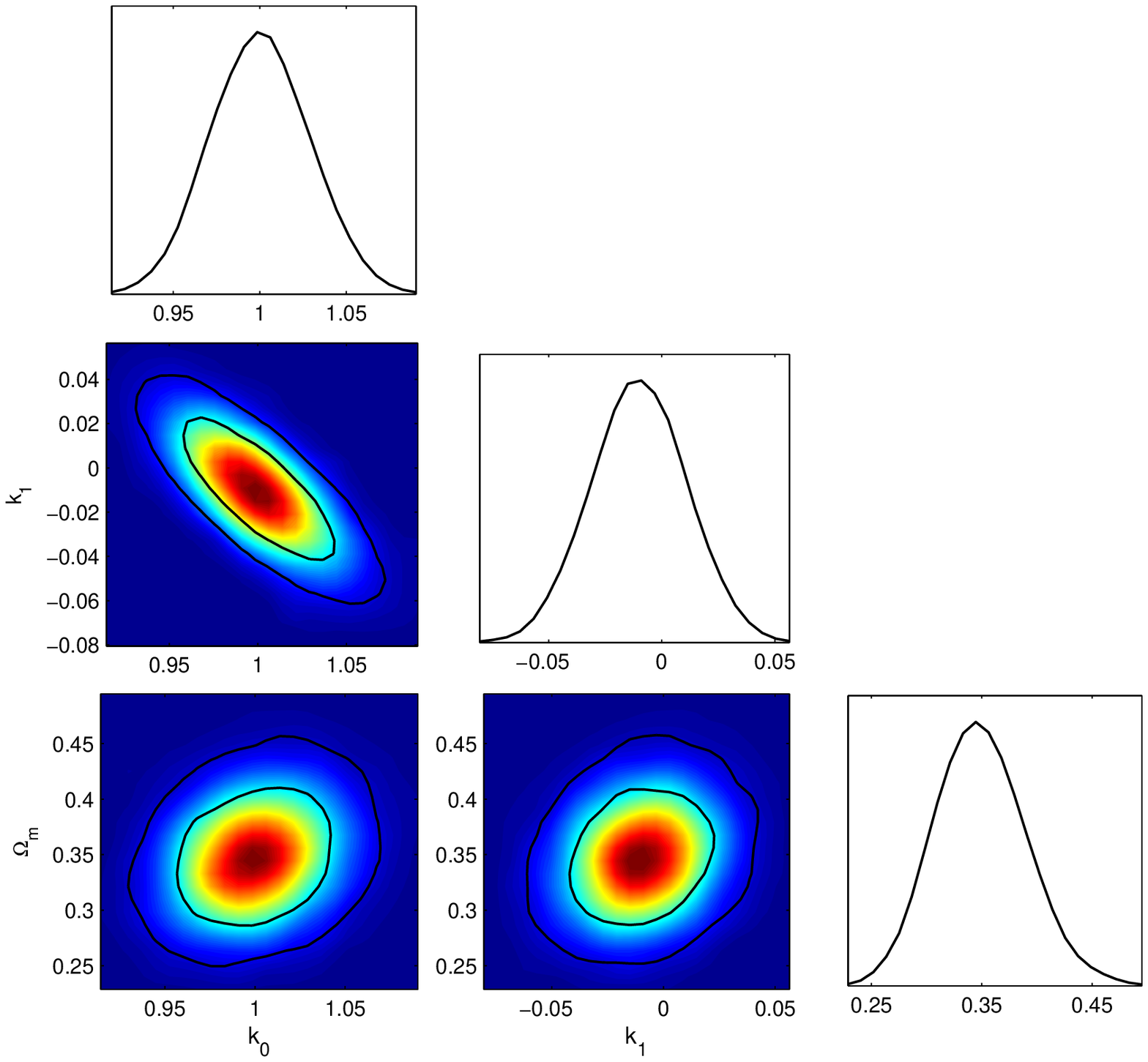}
  \end{center}
\caption{ Normalized posterior likelihood of $\Omega_m$ in the flat
$\Lambda$CDM model, which is derived from multi-frequency P15
sample.}\label{LCDM1}
\end{figure*}

\begin{figure*}
\begin{center}
\includegraphics[scale=0.45]{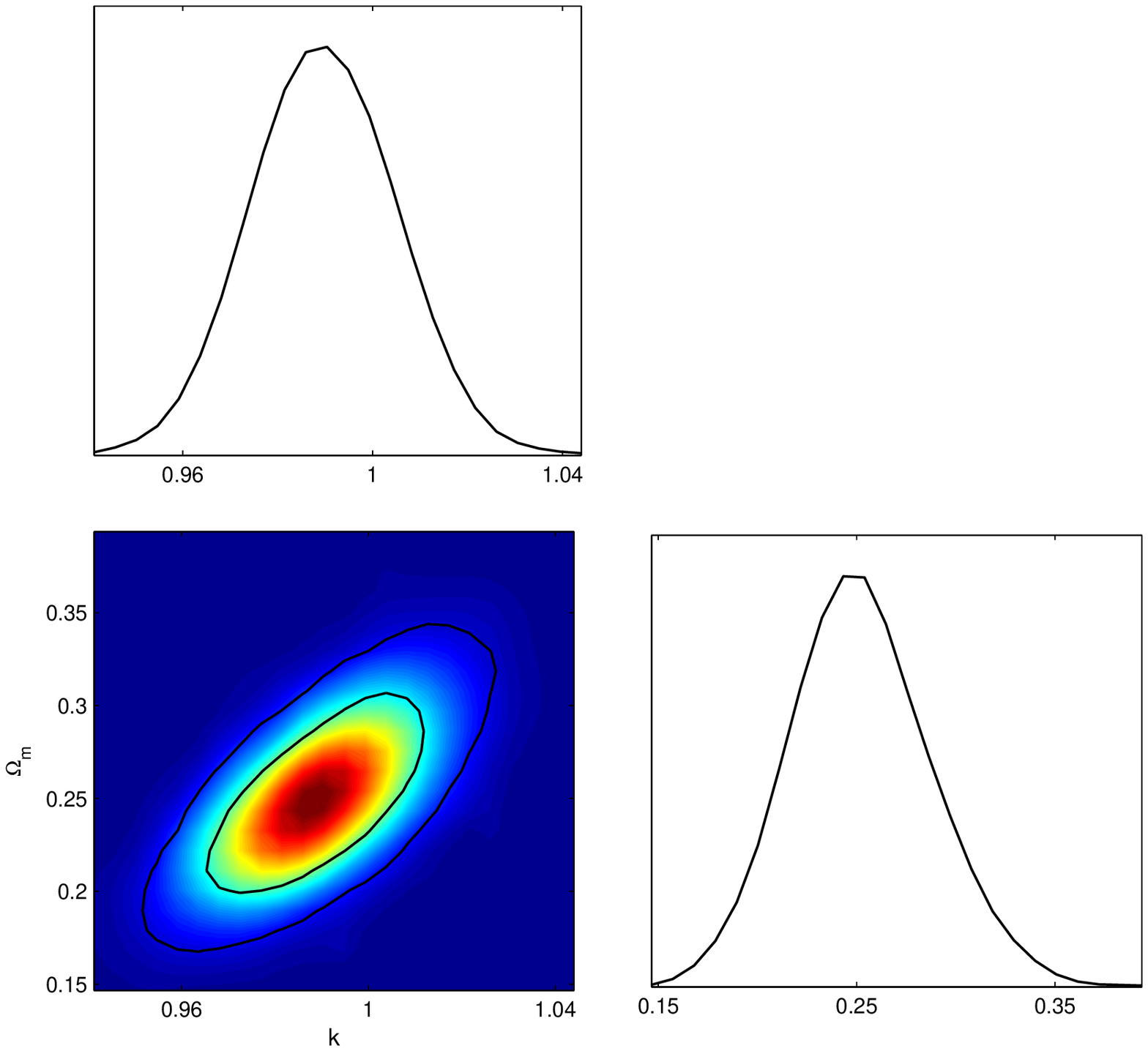} \includegraphics[scale=0.45]{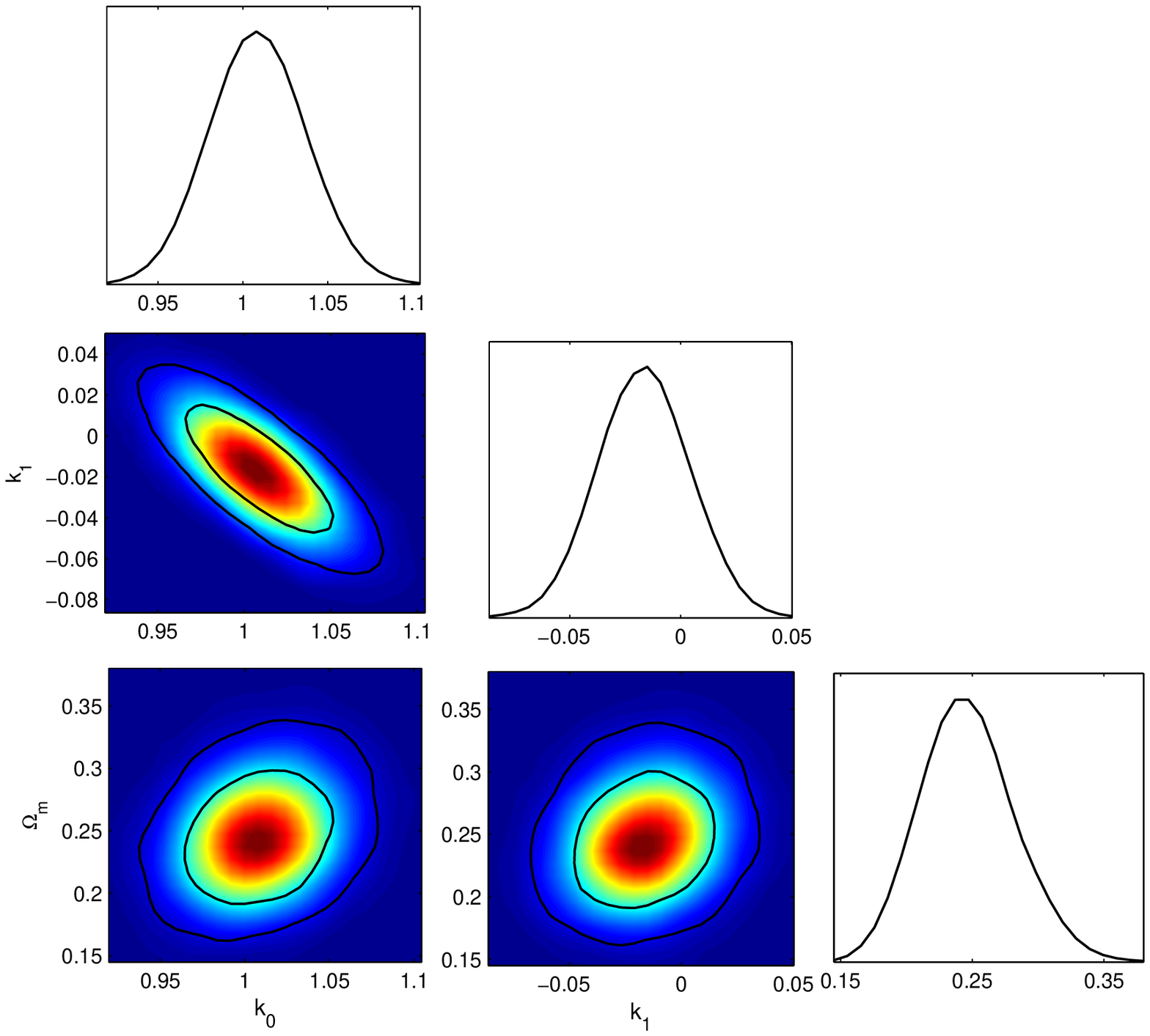}
  \end{center}
 \caption{Cosmological constraints on the flat DGP model from multi-frequency P15 sample. }\label{DGP}
\end{figure*}

\begin{figure*}
\begin{center}
\includegraphics[scale=0.45]{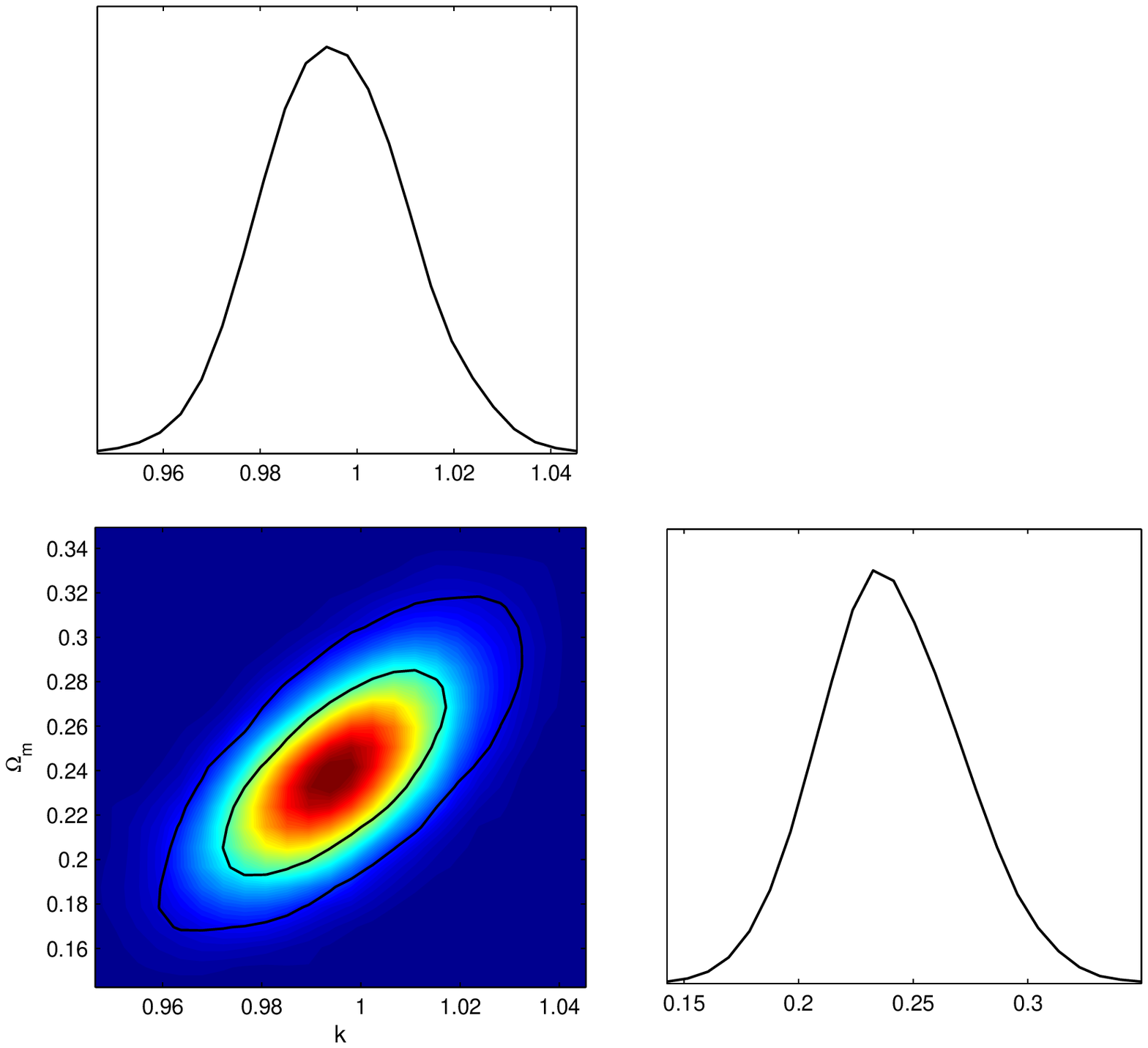} \includegraphics[scale=0.45]{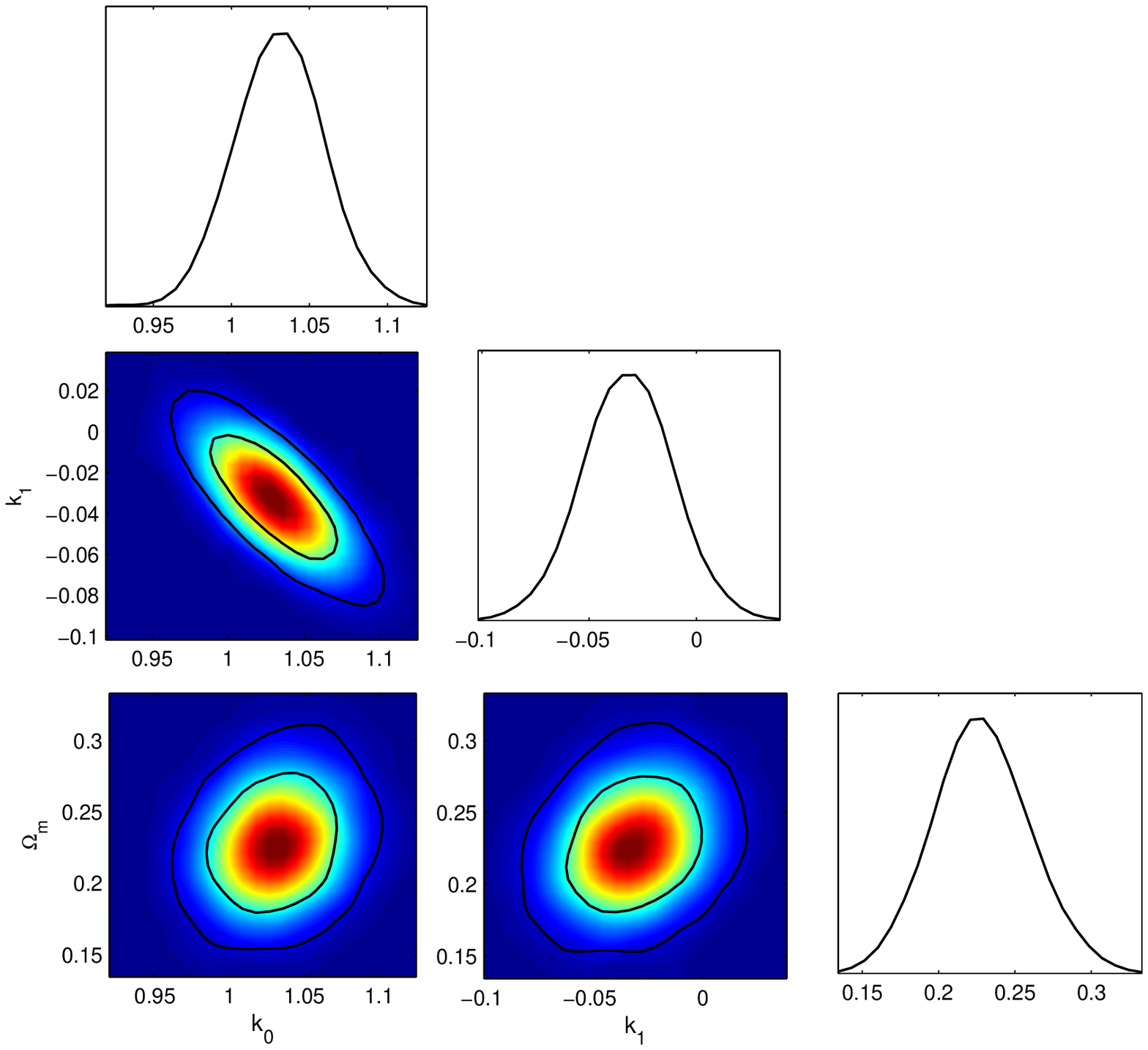}
  \end{center}
\caption{ Normalized posterior likelihood of $\Omega_m$ in the flat
$\Lambda$CDM model with Hubble constant from Riess
(2016).}\label{LCDM2}
\end{figure*}

\section{Results and discussion}

In order to demonstrate how the above described approach works we
have constrained two simple cosmological models using the
multi-frequency quasar data from Table~1. The models we chose are:
the $\Lambda$CDM and Dvali-Gabadadze-Porrati (DGP) models under
assumption of spatially flat Universe. In our fits the parameter $k$
representing the dependence of the angular size on frequency and its
evolution with redshift are fitted together with cosmological
parameters.

In our analysis we assumed flatness of the FRW metric, which is
strongly indicated by the location of the first acoustic peak in the
CMBR \citep{Ade14}. This conclusion is also independently supported
by the quasar data at $z\sim 3.0$ as demonstrated in \citep{Cao17a}.
It is a well known fact for the $\Lambda$CDM model while the DGP
model requires a few words of reminder. This model is one of the
simplest modified gravity models based on the concept of braneworld
theory \citep{Dvali00}, in which gravity leaks out into the bulk
above a certain cosmological scale $r_c$. Hence this scale is a free
parameter of the theory which in the flat DGP model can be
associated with the density parameter:
$\Omega_{r_c}=1/(4r^2_cH^2_0)$. It is then easy to see that the
relation $\Omega_{r_c} = \frac{1}{4} (1 - \Omega_m)^2$ is valid. The
results for different cosmological scenarios on the multi-frequency
VLBI observations are listed in Table 1 and discussed in turn in the
following sub-sections. The marginalized probability distribution of
each parameter and the marginalized 2D confidence contours are
presented in Fig.~3-6. In addition, we add the prior for the Hubble
constant $H_{0}=67.3$ km $\rm s^{-1}$ $\rm Mpc^{-1}$ based on the
recent Planck observations \citep{Ade14}.

We started our analysis with \textbf{the} $\Lambda$CDM model with
constant dark energy density and constant cosmic equation of state
$w=p/\rho=-1$, while two cases of conical jet model were considered:
a non-evolving constant $k$ parameter and an evolving one (denoted
in Table 2 as $\Lambda$CDM1 and $\Lambda$CDM2, respectively).
Parameters $k$, $k_0$ and $k_1$ were treated as free parameters to
be fitted.

Firstly, based on the prior for the Hubble constant after Planck
Collaboration XVI (2014), the likelihood is maximized at
$\Omega_m=0.353\pm0.063$ and $k=0.988\pm0.025$ with multi-frequency
measurements. In the second case, when the $k$ parameter
representing the dependence of the angular size on frequency is
allowed to evolve: $k(z)=k_0+k_1z$, treating intermediate luminosity
quasars as standard rulers, whose intrinsic length determined at 2
GHz scales as $l_m\propto\nu^{-k}$ to any other frequency, the
best-fit values for the parameters are $\Omega_m=0.348\pm0.063$ and
$k_0=0.999\pm0.043$, and $k_1=-0.009\pm0.0321$. The results are
illustrated in Fig.~3. One can easily see that multi-frequency VLBI
observations lead to reasonable cosmological fits, which motivates
us to improve constraints with a larger quasar sample from future
VLBI observations based on better uv-coverage. These results are
illustrated in Fig.~\ref{LCDM1} and Table~\ref{result}. For
comparison, one should refer to recent results obtained with other
independent precise measurements. First of all, the best fit value
of matter density parameter in the framework of flat $\Lambda$CDM
model reported by the \textit{Planck} collaboration \citep{Ade14}
was ${\Omega_m}=0.315\pm0.017$. Then, \citet{Hinshaw13} gave the
best-fit parameter $\Omega_m=0.279$ for the flat $\Lambda$CDM model
from the WMAP 9-year results, while the Hubble constant was
constrained by them at $H_{0}=70.0\pm2.2$ km $\rm s^{-1}$ $\rm
Mpc^{-1}$. One can see that our results are consistent with these
estimates. One can consider it as a consistency between the same
type of probes -- standard rulers -- acoustic peaks revealed by CMBR
anisotropy measurements at the redshift of $z \sim 1000$ and our
quasar sample reaching the redshift of $z \sim 3.0$. One can see
that $k$ parameter obtained from the multi-frequency sample is
consistent with the \citet{Blandford79} conical jet model ($k=1$) at
1$\sigma$ confidence level. Fits on the $k$ parameter also reveal
compatibility between our sample of quasars and the recent
multi-frequency core shift measurements of many compact radio
sources: Quasar 3C 345 \citep{Lobanov98} with extensive long-term
VLBI monitoring database at four frequencies (5, 8.4, 10.6, 22.2
GHz); Quasar 0850+581 \citep{Kovalev08} with a dedicated VLBA
experiment at 5, 8, 15, 24, and 43 GHz; Quasar 3C 309.1 and BL Lac
object 0716+714 \citep{Kovalev09} observed with VLBA at four
frequencies (8.1, 8.4, 12.1, 15.4 GHz); Blazar 1418+546, 2007+777,
and 2200+420 \citep{O'Sullivan09} observed with VLBA at eight
frequencies (4.6, 5.1, 7.9, 8.9, 12.9, 15.4, 22.2, 43.1 GHz).

Concerning the DGP model, marginalized distribution of the model
parameters are shown in Fig.~\ref{DGP}. Considering the case that
both $\Omega_m$ and $k$ are free parameters (denoted in Table 2 as
DGP1), we get the marginalized 1$\sigma$ constraints of the
parameters $\Omega_m=0.252\pm0.055$ and $k=0.989\pm0.024$. Working
on the evolving $k$ parameter representing the dependence of the
angular size on frequency, we find that the mass density parameter
in DGP model ($\Omega_m=0.245\pm0.054$) agrees very well with the
respective value derived from joint analysis of standard rulers and
standard candles including the measurements of the baryon acoustic
oscillation (BAO), cosmic microwave background (CMB), strong
gravitational lensing (SGL) and Type Ia supernovae (SNe Ia):
${\Omega_m}=0.267\pm0.013$ \citep{Biesiada11}. Constraints on the
redshift-evolving $k$ parameters ($k_0=1.009\pm0.043$,
$k_1=-0.016\pm0.032$) are even more consistent with the
\citet{Blandford79} conical jet model and previous analysis
especially with multi-frequency core shift measurements.

Finally, in the previous analysis we have assumed \textbf{the} prior
for the Hubble constant after Planck Collaboration XVI (2014), a
local determination of $H_{0}=73.24\pm1.74$ km $\rm s^{-1}$ $\rm
Mpc^{-1}$ with $2.4\%$ uncertainty from \citet{Riess16} can be taken
to perform consistency test. Such choice enables us to see the
influence of the Hubble constant on the constraining power of our
multi-frequency quasar data. Note that the Hubble constant derived
from the GP-processed $H(z)$ measurements in our analysis,
$H_{0}=69.2\pm3.7$ km $\rm s^{-1}$ $\rm Mpc^{-1}$, is well
consistent with the above three priors on $H_0$ at 68.3\% confidence
level. Such consistency has been extensively discussed in many
previous works focusing on improved constraints on the Hubble
constant through different model-independent methods
\citep{LiZ16,Wang16}. Here, we estimate the constraint results of
the flat $\Lambda$CDM with constant and evolving $k$ parameter,
which are specifically shown in Fig.~\ref{LCDM2} and
Table~\ref{result}. It is obvious that the values of matter density
obtained with the prior on $H_0$ taken after \citet{Riess16},
$\Omega_m=0.239\pm0.047$ (with $k_=0.995\pm0.023$) and
$\Omega_m=0.229\pm0.050$ (with $k_0=1.030\pm0.042$,
$k_1=-0.031\pm0.032$), are generally lower than that given by most
of other types of cosmological observations. This illustrates the
importance of measuring the Hubble constant accurately with
independent techniques and better understanding the nature of
discrepancy between $H_0$ inferred from CMBR or BAO and from local
measurements based on cosmic distance ladder. Summarizing, although
the cosmological constraints become much weaker with the inclusion
of different systematics, the current standard cosmological model
($\Omega_m \sim 0.3$) with a significant cosmological constant
($\Omega_\Lambda \sim 0.7$) in the flat universe is still preferred
by our quasar sample at high confidence level.

\section{Conclusions} \label{sec:conclusions}

In conclusion, our analysis demonstrates that multi-frequency
angular size measurements of intermediate-luminosity quasars
reaching the redshifts $z\sim 3$ can be used as standard rulers for
cosmological inference. Therefore, one may say that the approach
initiated in \citet{Cao16,Cao17a} can be further developed. More
importantly, our results indicate that, the multi-frequency quasar
sample is consistent with the \citet{Blandford79} conical jet model
($k=1$), which, from the physical point of view means that opacity
of the jet is governed by pure synchrotron self-absorption, i.e.
external absorption does not play any significant role in the
observed angular sizes at least up to 43 GHz. One should stress that
the present paper is only the first step toward elaborating the
scheme to identify and calibrate compact radio-sources as standard
rulers taking advantage of multi-frequency observations. Still,
there are several remarks that remain to be clarified as follows.

Firstly, in performing the statistical analysis from the
multi-frequency quasar data, we find that a precise determination of
the linear size through a cosmological model-independent method
plays an significant part to achieve such goal, which remains to be
addressed in future analyses on a larger sample. On the one hand, we
demonstrated that the approach initiated in \citet{Cao17a}, i.e.,
calibrating intermediate luminosity milliarcseconds compact radio
quasars in sufficiently big sample obtained even at a single
frequency is promising. Namely, the intrinsic metric size $l_m$
identified at some observing frequency, when properly rescaled, can
be used in objects observed in other surveys performed at other
frequencies. On the other hand, although the cosmological
constraints derived from multi-frequency data in the current study
agrees very well with that form the previous sample based on
single-frequency data, this inference still heavily relies on the
assumption of \citet{Blandford79} conical jet model, i.e., the radio
core tends to be self-absorbed and in equipartition. Moreover, it is
of paramount importance to determine precisely to which extent this
holds for the intermediate luminosity quasars, which should be done
in the future on a much larger sample.

Secondly, there are several sources of systematics we do not
consider in this paper and which remain to be addressed in the
future analysis. Let us start with the possible biases associated
with sample incompleteness. One general concern is given by the fact
that there are different ranges of frequencies addressed for each
source in the compiled data by \citet{Pushkarev15}. Although this
problem has been recognized long time ago, the most straightforward
solution to this issue is focusing on a larger sample of
multi-frequency measurements of compact structure covering the same
range of frequencies, which is hard to be rigorously accounted in
context of cosmological studies like in this paper. The other
systematic is the scattering in the size determination of objects at
cosmological distances, which would furthermore affect the derived
value of the $k$ parameter. More specifically, if the intergalactic
medium broadens the signal of the radio sources, the apparent size
will be larger at larger distances \citep{Gurvits94,Gurvits99}.
Further progress has recently been achieved in the study of Milky
Way scattering properties and intrinsic sizes of active galactic
nuclei cores \citep{Pushkarev15}. The so-called Galactic broadening
is found to be dependent on the galactic latitude and intrinsic
scattering, while its strength is possibly correlated with the
Galactic H$\alpha$ intensity, free-electron density, and Galactic
rotation measurements. However, one should note that such effect is
still difficult to be precisely quantified, especially in the case
when the Universe is filled with charged particles.

Thirdly, even with a relatively small sample of 30 sources we were
able to demonstrate that combined multi-frequency data concerning
compact radio quasars gives quite stringent cosmographic constraints
and is able to differentiate between different cosmological models
like $\Lambda$CDM and DGP. The value of density parameter in
$\Lambda$CDM model is perfectly consistent with values obtained in
an independent manner. Moreover when confronted with alternative
methods of determining $\Omega_m$ like from the peculiar velocities
of galaxies \cite{Feldman2003} our fits obtained for the DGP model
are accurate enough to falsify this model. However, strong
degeneracy between $H_0$ and $\Omega_m$, illustrated in our study in
the spirit of sensitivity analysis, emphasizes the importance of
independent and more direct determinations of $H_0$. In this respect
the approach of strong lensing time delays is promising. Recent
results of the H0LiCOW project \citep{Holicow} already demonstrated
that a few percent accuracy in $H_0$ determination is feasible.

Finally, the results presented in this paper pave the way for the
follow up engaging multi-frequency VLBI observations of more compact
radio quasars with higher sensitivity and angular resolution, which
may make it less susceptible to systematic errors. The approach,
introduced in this paper, would make it possible to build a
significantly larger sample of standard rulers at much higher
redshifts. With such a sample, we can further investigate
constraints on the cosmic evolution and eventually probe the
evidence for dynamical dark energy \citep{Wei16,Cao17c}.

\section*{Acknowledgments}

We are grateful to John Jackson for useful discussions. This work
was supported by the National Key Research and Development Program
of China under Grants No. 2017YFA0402603; the Ministry of Science
and Technology National Basic Science Program (Project 973) under
Grants No. 2014CB845806; the National Natural Science Foundation of
China under Grants Nos. 11503001, 11373014, and 11690023; Beijing
Talents Fund of Organization Department of Beijing Municipal
Committee of the CPC; the Fundamental Research Funds for the Central
Universities and Scientific Research Foundation of Beijing Normal
University; China Postdoctoral Science Foundation under grant No.
2015T80052; and the Opening Project of Key Laboratory of
Computational Astrophysics, National Astronomical Observatories,
Chinese Academy of Sciences. This research was also partly supported
by the Poland-China Scientific \& Technological Cooperation
Committee Project No. 35-4. M.B. was supported by Foreign Talent
Introducing Project and Special Fund Support of Foreign Knowledge
Introducing Project in China.

\end{document}